# OSCILLATORY THICKNESS DEPENDENCE OF THE COERCIVE FIELD IN MAGNETIC 3D ANTI-DOT ARRAYS

A. A. Zhukov[1], M. A. Ghanem[2], A. V. Goncharov[1], R. Boardman[3], V. Novosad[4], G. Karapetrov[4], H. Fangohr[3], P. N. Bartlett[2] and P. A. J. de Groot[1]

[1]School of Physics and Astronomy, [2]School of Chemistry, [3]School of Engineering Sciences, University of Southampton, Southampton, SO17 1BJ, UK, [4]Materials Science Division, Argonne National Laboratory, 9700 South Cass Ave., Argonne, IL60439, USA

We present studies on magnetic nano-structures with 3D architectures, fabricated using electrodeposition in the pores of well-ordered templates prepared by self-assembly of polystyrene latex spheres. The coercive field is found to demonstrate an oscillatory dependence on film thickness reflecting the patterning transverse to the film plane. Our results demonstrate that 3D patterned magnetic materials are prototypes of a new class of *geometrical multilayer structures* in which the layering is due to local shape effects rather then compositional differences.

The behaviour of magnetic particles with sub-micrometer sizes has attracted the attention of researchers for many years. Until recently only disordered structures were available[1]. From the early work on such structures the foundations of micro-magnetism were established[2-4]. The development of lithographic techniques has allowed researchers to prepare ordered sub-micron structures[5-9]. Often these are two-dimensional arrays of small magnetic elements (dots) or arrays of holes in magnetic films (anti-dots).

Advances in fabrication have renewed the interest in nano-magnetism, in particular in the mechanisms governing magnetization reversal and in the behaviour of structures for which the dimensions of the magnetic element is of the order of characteristic length scales of the magnetic material (*e.g.* domain wall width, exchange length). In the traditional lithographic approach low sub-micron length scales are usually achieved by means of electron or particle beam lithography, which is slow and costly. To realize the potential of sub-micron magnetics, alternative fabrication methods have to be developed. Self-assembly methods, such as the one presented in this work, offer promising preparation routes[10-13].

Fabrication methods based on templates formed by the self-assembly of colloidal particles have been considered for various applications such as photonic materials[14-16], microchip reactors[17] and biosensors[18]. In this Letter we describe studies on nano-scale magnetic materials prepared by electrochemical deposition in ordered templates made from polystyrene latex spheres with diameters ranging from 50 nm to 1000 nm[19, 20]. This technique offers new opportunities, which are not easily realized by standard lithographic methods, and allows us to create magnetic nano-structures with 3D architectures on a broad range of length scales.

The coercive field $B_c$ is determined by a combination of intrinsic properties of the magnetic material and by magnetic dipolar interactions. The latter give rise to shape effects. There has been much interest in the dependence of $B_c$ on film thickness. In contrast to non-patterned thin films, for which the coercive force is known to increase monotonically with decreasing thickness[1, 21], we have found that $B_c$ for our 3D nanostructured films reveals a novel oscillatory behaviour.

The crucial element of our technique is a well-ordered template of mono-disperse latex spheres. We deposit the template on a glass substrate, which has buffer layers of Cr (10 nm) and Au (200 nm) deposited by sputtering. The template has been prepared using a slow (3-5 days) evaporation of a colloidal water suspension containing 0.5 wt% of latex spheres[20, 22, 23]. The capillary forces combined with the electrostatic repulsion between the spheres create a well-ordered close packed structure. Using the template as a mould we prepare nano-porous magnetic structures by electro-deposition from the gold underlayer[20, 22, 23]. After deposition the latex spheres can be removed by dissolving in toluene. Using this method, well-ordered nano-structured 3D anti-dot arrays were prepared for various magnetic materials such as cobalt, iron, nickel and soft-magnetic $Ni_{50}Fe_{50}$ alloy. The array period and the film thickness were controlled by changing the diameter, $d$, of the polystyrene latex spheres, used to form the template, and the amount of charge passed during the electro-deposition process respectively[22]. The morphology of the films, their composition and crystal structure were characterized using scanning electron microscopy (SEM), energy dispersive X-ray spectroscopy (EDXRS) and X-ray diffraction. This confirmed that the electrodeposition results in films with a homogeneous composition and with excellent filling of the template voids. After the removal of the latex spheres the structure was found to be very stable without any detectable shrinkage or cracking.

SEM images of the nano-structured films reveal excellent hexagonal order (Fig. 1). The size of ordered domains reaches 1mm and an average concentration of point defects in the anti-dot lattice of 0.1%. In comparison with standard lithographical techniques our template deposition method has a significant advantage. It allows us to create patterning in the direction transverse to the film plane. Cross-sectional SEM (Fig.1b) shows that the transverse structuring is also well ordered. Combined with the broad range of accessible length scales, this offers promise for the preparation of nano-structured magnetic materials such as data storage media and elements for magneto-electronics sensitive to resonant electromagnetic or acoustic waves.

The nano-structuring significantly affects the shape of magnetization loops and drastically changes the coercive field $B_c$ which we have found to show a maximum with variation of sphere diameter[24], resembling that of particulate magnetic materials[25]. Measurements of the dependence of $B_c$ on the thickness of the magnetic film, $t_f$, revealed a remarkable behaviour. Although these films have a homogeneous composition of magnetic material, as evident from the EDSRX studies, we have found that for all materials investigated the coercive force changes periodically with film thickness. This is a clear manifestation of the structuring in the direction *transverse* to the film and the 3D architecture of these structures. Fig.2 demonstrates that $B_c$ shows pronounced oscillations and reaches a maximum for the case when the top surface of the film is near the centre of a layer of close packed spherical voids. For complete spherical layers the coercive field approaches a minimum. These observations suggest that the points where the spheres touch play an important role in increasing local anisotropy and hence the coercivity.

To investigate the local magnetic structures of the templated films, we employed magnetic force microscopy (MFM) complemented by micromagnetic modelling. MFM images were obtained with a Digital Instruments 3000 scanning probe microscope using lift-mode with a fly height of 100 nm and a standard low-moment ferromagnetic tip. Results are shown in Fig. 3a for a relatively thin (100 nm) Co film with $d = 700$ nm in the remanent state. These images reveal ordered, rhombic magnetic patterns associated with the hole array. Magnetic structures for 2D hole arrays were calculated using micromagnetic modelling with the OOMMF software suite[26]. Parameters in the computation were chosen in accordance with the known values for Co[25] and hole diameters of 0.4 $d$. The micromagnetic simulations reveal that, to decrease total magnetic energy, the magnetic structure divides up into three different domains: two in regions between nearest neighbour holes and the third a diamond-shape bounded by four holes (Fig. 3b). This breaks the triangular symmetry of the array. Two out of the three regions between holes contain domains with their magnetizations aligned perpendicular to the vector connecting the nearest neighbour hole centres. The remaining third is part of the diamond-shape and has a magnetization aligned parallel to the vector connecting hole centres. Comparing these magnetic structures with those found for square arrays[27], indicates a universal topology for regular anti-dot arrays.

Furthermore 2D micromagnetic simulations have been used to analyse the magnetisation of hexagonal arrays of holes with period $d$ and diameter $D$. The parameters for the micromagnetic modeling of $Ni_{0.5}Fe_{0.5}$ were chosen to be $A = 5.85$ pJ/m (exchange), $\mu_o M_S = 1.58$ T (saturation magnetization) and $K = 0$ (anisotropy). As can be seen from Fig.4a, $B_c$ increases strongly when $D$ approaches the array period $d$. This originates from a large increase in local anisotropy energy when the regions between holes become constricted. This observation can be used to explain qualitatively the experimentally observed oscillatory behaviour $B_c(t_f)$ in the 3D structure assuming that with variation of thickness we periodically add soft and hard magnetic layers depending on the $D/d$ ratio. To explain the $B_c(t_f)$ data in a more quantitative manner would require full 3D micromagnetic modelling. However due to the large number of computational cells required to deal correctly with the range of relevant length scales – from several sphere diameters (~1 μm) down to the exchange length (*e.g.* ~ 2 nm for Co[25]) – and the 3D nature of these phenomena, this is a gargantuan task. Therefore we have employed a simplified approach. We consider our system as a multilayer with properties averaged in the plane direction. Using the values of $B_c(D/d)$ found from the 2D numerical simulations discussed above, we model a 1D stack of layers with the anisotropy parameter for a given layer: $K(D/d) = m_S B_c(D/d) /18$, with $m_S$ the saturation magnetic moment of this layer. Using a Monte-Carlo method we calculate $B_c(t_f)$ for this stack of exchange coupled layers. This method is similar to calculations of the coercivity in hard-soft exchange spring multilayer systems[28]. As can be seen from Fig.4b the results of the numerical simulations reproduce our experimental behaviour. Note that for simplicity we have chosen the period in the perpendicular direction to be $d$ in this model; the period of the $B_c(t_f)$ oscillations reflects this.

In an alternative approach, we have described the behaviour of 3D anti-dot arrays also in terms of a simple domain wall model where wall pinning is the main coercivity mechanism. In this model, which is similar to well-established models of domain wall pinning by inclusions[29], we assume a flat domain wall with width δ. The domain wall energy is taken proportional to the domain wall volume and is calculated as function of the position of the wall in the f.c.c. anti-dot structure. This allows the calculation of the coercive force which reproduces well the $B_c(t_f)$ oscillations found in our measurements (Fig. 4b).

The studies, presented in this Letter, of 3D magnetic nanostructures obtained from sphere templates reveal the presence of layers with increased pinning characteristics. As a consequence they resemble multilayers with alternating hard and soft magnetic layers. In contrast to conventional multilayer systems where the changes in layer properties result from compositional differences, in the 3D nanostructured films the differences are fundamentally due to geometry. Hence these materials constitute a new class of *geometrical multilayer structures* and present promising directions for further research.

This work was supported by the Engineering and Physical Sciences Research Council (UK) under grant GR/N29396.

**Figure Captions**

Fig.1. SEM image of structured Ni film prepared using 0.5 µm polystyrene spheres (A) and cross-sectional view of a thick film after cleaving (B). In both cases the white marker corresponds to 1 µm.

Fig.2. Coercive fields for $Ni_{50}Fe_{50}$ films with different values of thickness, $t_f$. The dashed lines indicate the positions of sphere centers for each layer in the close-packed structure. $B_c$ shows maxima close to the sphere centers. The solid line is a guide to the eye.

Fig.3. MFM image – measured for a Co anti-dot film with $d = 700$ nm and $t_f = 100$ nm (a). Blue circles indicate the hole positions. The image on the right (b) gives the magnetic structure in the film obtained from micromagnetic modeling. The color bar gives the angle of the magnetization direction in radians.

Fig.4. Coercive field of films with hexagonal hole arrays from 2D numerical simulation as a function of the hole diameter $D$ for fixed period of $d = 0.5$µm (a). On the right (b) the calculated dependence of the coercive field from the homogeneous-layer micromagnetic model and from the domain wall pinning model (with $d/\delta = 25$).

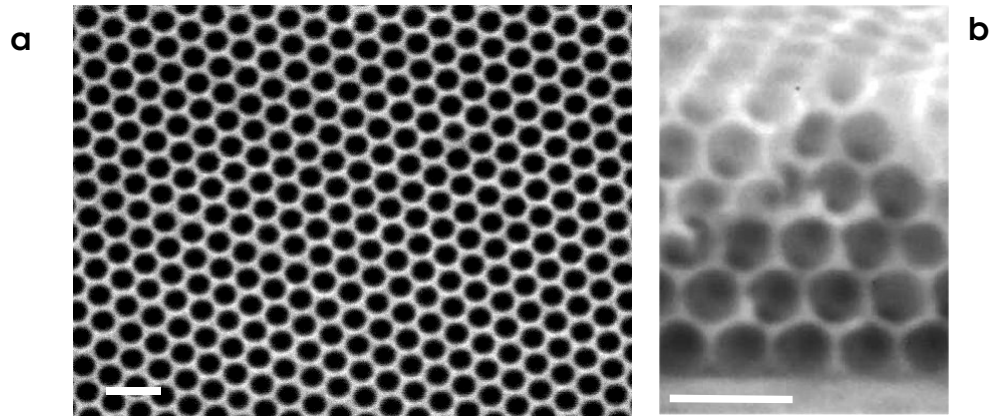

a
b

**Fig. 1**

A.A.Zhukov et al., "OSCILLATORY THICKNESS DEPENDENCE OF THE

COERCIVE FIELD IN MAGNETIC 3D ANTI-DOT ARRAYS"

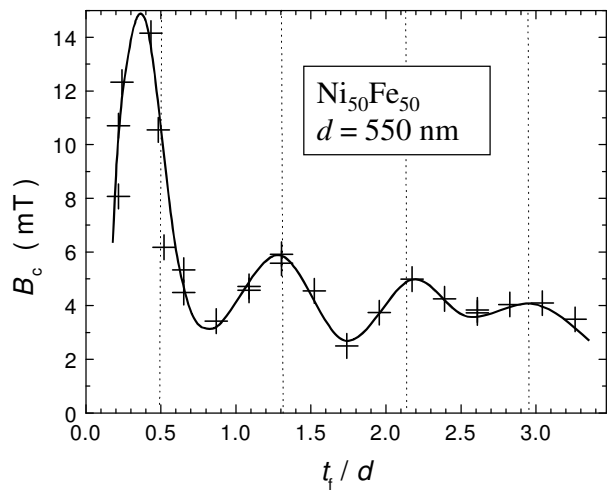

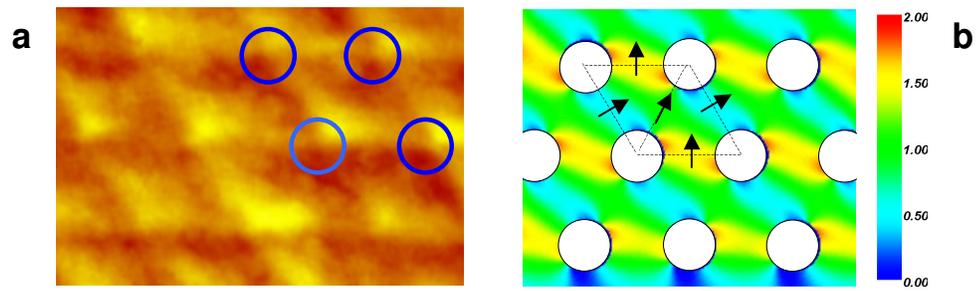

Fig.3

A.A.Zhukov et al., "OSCILLATORY THICKNESS DEPENDENCE OF THE COERCIVE FIELD IN MAGNETIC 3D ANTI-DOT ARRAYS"

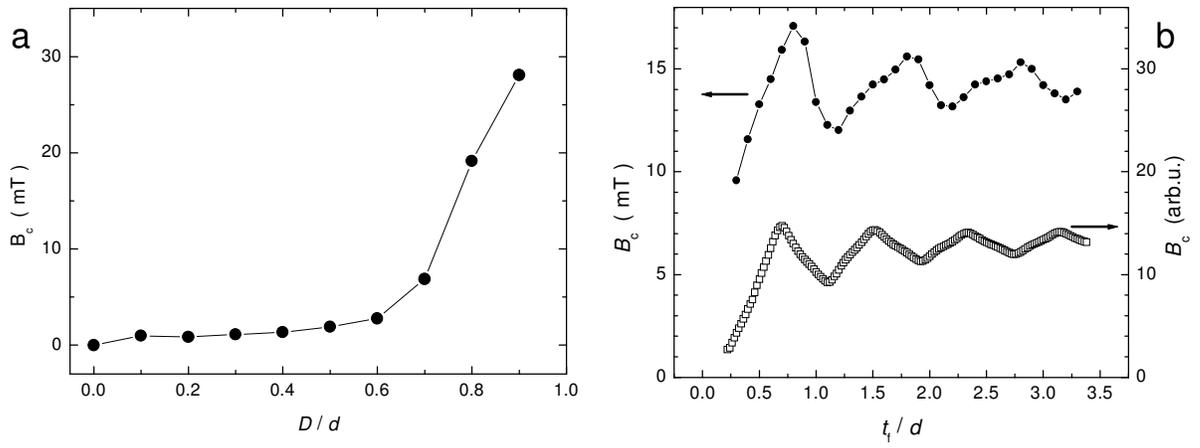

Fig. 4

A.A.Zhukov et al., "OSCILLATORY THICKNESS DEPENDENCE OF THE

COERCIVE FIELD IN MAGNETIC 3D ANTI-DOT ARRAYS"